\documentclass[amsmath,amssymb,nofootinbib,onecolumn]{revtex4-2}

\usepackage{graphicx}
\usepackage{dcolumn}
\usepackage{bm}

\usepackage{hhline}
\usepackage[american]{babel}
\usepackage{booktabs}
\usepackage{dcolumn}  
\usepackage[T1]{fontenc}  
\usepackage[colorlinks=true,
            linkcolor=red,     
            citecolor=blue,      
            urlcolor=blue]{hyperref}  
\usepackage[utf8]{inputenc}  
\usepackage{multirow}
\usepackage{times}  
\usepackage{url}
\usepackage[dvipsnames]{xcolor}
\usepackage{float}
\usepackage{resizegather}
\usepackage{appendix}

\newcolumntype{d}[1]{D{.}{.}{#1}}
\newcolumntype{v}[1]{D{,}{,\ }{#1}}

\begin{document}
	
	
	\title{Higher-order gravity models: corrections up to cubic curvature invariants and inflation}

    \author{C. M. G. R. Morais}
	\email{caio.morais.113@ufrn.edu.br}
	\affiliation{Departamento de Física, Universidade Federal do Rio Grande do Norte,\\
	Campus Universitário, s/n - Lagoa Nova, CEP 59072-970, Natal, Rio Grande do Norte, Brazil}
    
	\author{G. Rodrigues-da-Silva}
	\email{gesiel.neto.090@ufrn.edu.br}
	\affiliation{Departamento de Física, Universidade Federal do Rio Grande do Norte,\\
	Campus Universitário, s/n - Lagoa Nova, CEP 59072-970, Natal, Rio Grande do Norte, Brazil}
	
	\author{L. G. Medeiros}
	\email{leo.medeiros@ufrn.br}
	\affiliation{Escola de Ciências e Tecnologia, 	Universidade Federal do Rio Grande do Norte,\\
	Campus Universitário, s/n - Lagoa Nova, CEP 59072-970, Natal, Rio Grande do Norte, Brazil}

	\date{\today}

\begin{abstract}
\vspace{0.2 mm}
We construct a higher-order gravity model including all corrections up to mass dimension six. Starting from the Jordan frame, we derive the field equations and specialize to the FLRW background, where the dynamics take the form of a four-dimensional autonomous system. Focusing on the $R+R^{2}+RR_{\mu\nu}R^{\mu\nu}$ case, we obtain linearized equations in the parameter $\gamma_{0}$ and analyze the resulting phase space. The model exhibits the main desirable features of an inflationary regime, with a slow-roll attractor and a stable critical point corresponding to the end of inflation. Analytical expressions for the scalar spectral index $n_{s}$ and the tensor-to-scalar ratio $r$ show that the model is consistent with Planck, BICEP/Keck, and BAO data if $|\gamma_{0}|\lesssim 10^{-3}$. Moreover, negative values of $\gamma_{0}$ restore compatibility with recent ACT, Planck, and DESI results, suggesting that higher-order corrections may be relevant in refining inflationary cosmology.

\end{abstract}

\maketitle

\section{Introduction}

From a theoretical perspective, General Relativity (GR) can be derived from
Lovelock's theorem \cite{Lovelock1971}. Developed by David Lovelock in the early
1970s, this theorem states that "\textit{the only second-order differential
equation that can be derived from an action depending solely on the metric in
four dimensions is Einstein's equation}" \cite{Clifton2012}. More specifically,
Lovelock's theorem is built upon four assumptions: \emph{(a)} the integration
region of the action is four-dimensional; \emph{(b)} the metric is the only
field entering the action; \emph{(c)} the field equations are invariant under
general coordinate transformations; \emph{(d)} the resulting field equations
must be second-order differential equations. Modifications to GR can be achieved
by violating any of these hypotheses. For instance, if we relax the assumption
that the metric is the only fundamental field and includes an additional scalar
degree of freedom, we obtain Horndeski theories \cite{Horndeski1974}. If,
instead, we consider a Riemann–Cartan geometry containing an affine connection
with a non-vanishing antisymmetric component, torsion is incorporated into the
description of gravitation, leading to the Einstein–Cartan theories
\cite{Trautman2006}. On the other hand, by allowing the field equations of the
theory to be of higher than second order, we encounter higher-order gravity
theories \cite{EPJC2008,PRD2016}.

Higher-order gravity theories are constructed from invariants involving the
Riemann tensor and covariant derivatives, which, when added to the
Einstein–Hilbert (EH) action, yield equations of motion of order higher than
two. A convenient way to classify each term of the higher-order action is by
its mass dimension. In natural units, the covariant derivative $\nabla_{\mu}$
has mass dimension one, while the curvature tensor $R_{\mu\nu\alpha\beta}$ has
mass dimension two. For instance, the Ricci scalar in the EH action is a term
of mass dimension two. In contrast, the invariants $R^{2}$,
$R_{\mu\nu}R^{\mu\nu}$, $R_{\mu\nu\alpha\beta}R^{\mu\nu\alpha\beta}$, and
$\square R$ constitutes the complete set of all terms of mass dimension four.
Moreover, in Ref.~\cite{Fulling1992}, the authors showed that there are $17$
independent combinations involving terms of mass dimension six. Within the
framework of effective field theories \cite{Rocci2024}, this type of
classification allows one to interpret the higher-order terms as higher-energy
corrections to GR. Thus, the first-order corrections to the EH action involve
terms of mass dimension four, whereas second-order corrections correspond to
mass dimension six terms, and so forth.

This work aims to construct the higher-order action including all
possible invariants up to mass dimension six, and subsequently to study the
inflationary cosmology of this theory. We will show in the following sections
that, in the cosmological background, the only nontrivial first-order
correction term is $R^{2}$. Thus, the higher-order action that accounts for
terms up to mass dimension four reproduces the well-known Starobinsky inflation
\cite{Starobinsky1980,Starobinsky1979}. Furthermore, among the $17$ mass
dimension six terms associated with second-order corrections, only three terms
effectively contribute to the cosmological background field equations. Although
there exist different combinations of three terms that fully characterize the
second-order corrections, in this work, we will consider the invariants $R^{3}$,
$R\square R$, and $RR_{\mu\nu}R^{\mu\nu}$. Therefore, in the inflationary
context, the higher-order model under consideration comprises, in addition to
the EH term, the contributions from the invariants $R^{2}$, $R^{3}$,
$R\square R$, and $RR_{\mu\nu}R^{\mu\nu}$.

The Starobinsky model is a remarkable success and is currently one of the most
promising candidates for generating an inflationary regime. The main reason for
this success is that it is a single-parameter model that fits recent CMB
observations very well \cite{Planck2020,BICEP3}. Moreover, considering the
energy scale of $10^{15}$ GeV at which Starobinsky inflation takes place, the
$R^{2}$ term can be interpreted as the leading higher-energy correction to GR.
Within this effective theory context, the invariants $R^{3}$, $R\square R$, and
$RR_{\mu\nu}R^{\mu\nu}$ are naturally regarded as corrections to the
Starobinsky model. Inflationary models involving the $R^{2}$ term together with
$R^{3}$ and $R\square R$ have been explored in the literature both at the
background level
\cite{Maeda1990,Gottlober1990,Gottlober1991,Amendola1993,Lihoshi2011} and at
the perturbative level
\cite{Huang2014,Castellanos2018,CMP2019,Cheong2020,SSM2022,SM2023}. In this
work, we will focus on the study of the inflationary model
$R+R^{2}+RR_{\mu\nu}R^{\mu\nu}$. Considering the homogeneous and isotropic
(FLRW) background, we will discuss the phase space of the theory and the
existence of a slow-roll regime. Furthermore, assuming that the invariant
$RR_{\mu\nu}R^{\mu\nu}$ is of the same order of magnitude as the Starobinsky
term, the impact of this invariant will be analyzed in light of the
observational constraints on the $n_{s}\times r$ plane \cite{BICEP3}.

Throughout this paper, we adopt the following conventions: natural units, metric
signature $(+,-,-,-)$, $M_{pl}^{2}=8\pi G$, and
$\square\equiv\nabla_{\mu}\nabla^{\mu}$.

\section{Higher-order gravity model\label{sec - EFT}}

The first step in constructing the proposed model is to enumerate all the
independent scalar invariants that can be built from the Riemann tensor and the
covariant derivative. By exploiting the symmetries of the curvature tensor and
the Bianchi identities, one can show that there exist 4 independent scalars
of mass dimension four:
\begin{equation}
R^{2},\text{ \ }\square R,\text{ \ }R_{\mu\nu}R^{\mu\nu}\text{ \ and
\ }R_{\alpha\mu\beta\nu}R^{\alpha\mu\beta\nu}, \label{eq:scalar_order_4}%
\end{equation}
and 17 independent scalars of mass dimension six:
\begin{align}
&  R^{3},\ \ RR_{\mu\nu}R^{\mu\nu},\ \ R_{\mu\nu}R_{\text{ \ }\rho}^{\mu
}R^{\nu\rho},\ \ R_{\alpha\beta}R_{\mu\nu}R^{\alpha\mu\beta\nu},\ \ RR_{\alpha
\mu\beta\nu}R^{\alpha\mu\beta\nu},\nonumber\\
&  R_{\alpha\rho}R_{\text{ \ }\mu\beta\nu}^{\rho}R^{\alpha\mu\beta\nu
},\ \ R_{\alpha\mu\beta\nu}R_{\text{ \ \ \ }\rho\sigma}^{\alpha\mu}%
R^{\rho\sigma\beta\nu},\ \ R_{\alpha\mu\beta\nu}R_{\text{ }\rho\text{
\ \ }\sigma}^{\alpha\text{ \ }\beta}R^{\mu\rho\nu\sigma},\ \ \nabla_{\lambda
}R\nabla^{\lambda}R,\nonumber\\
&  \nabla_{\lambda}R_{\mu\nu}\nabla^{\lambda}R^{\mu\nu},\ \ \nabla_{\lambda
}R_{\mu\nu}\nabla^{\mu}R^{\lambda\nu},\ \ \nabla_{\lambda}R_{\alpha\mu\beta
\nu}\nabla^{\lambda}R^{\alpha\mu\beta\nu},\ \ R\square
R,\label{eq:scalar_order_6}\\
&  R^{\mu\nu}\nabla_{\mu}\nabla_{\nu}R,\ \ R_{\mu\nu}\square R^{\mu\nu
},\ \ R^{\alpha\mu\beta\nu}\nabla_{\mu}\nabla_{\nu}R_{\alpha\beta
}\text{\hspace{0.5cm}and\hspace{0.5cm}}\square\square R.\nonumber
\end{align}
For details see Ref.~\cite{Fulling1992}.

Next, we take into account that these 21 invariants appear inside an integral
action of dimension $n$, and therefore several of them vanish or are related to
each other \cite{Decanini2007}. For example,
\begin{align*}
\int_{\Omega}\square R\sqrt{|g|}d^{n}x  &  =\int_{\Omega}\square\square
R\sqrt{|g|}d^{n}x=0\text{,}\\
\int_{\Omega}\nabla^{\lambda}R^{\mu\nu}\nabla_{\lambda}R_{\mu\nu}\sqrt
{|g|}d^{n}x  &  =-\int_{\Omega}R^{\mu\nu}\square R_{\mu\nu}\sqrt{|g|}%
d^{n}x\text{.}%
\end{align*}
Here we are neglecting surface terms. In this situation, the independent terms
of mass dimension four and six reduce to 3 and 10, respectively. Finally, by
considering $n=4$, we still have three constraints arising from Xu's geometric
identity \cite{Xu1987} and from Lovelock invariants \cite{Lovelock1971}:
\[
R^{3}-8RR_{\mu\nu}R^{\mu\nu}+8R_{\mu\nu}R_{\text{ \ }\rho}^{\mu}R^{\nu\rho
}+8R_{\alpha\beta}R_{\mu\nu}R^{\alpha\mu\beta\nu}+RR_{\alpha\mu\beta\nu
}R^{\alpha\mu\beta\nu}-4R_{\alpha\rho}R_{\text{ \ }\mu\beta\nu}^{\rho
}R^{\alpha\mu\beta\nu}=0\text{,}%
\]
and
\begin{align*}
\mathcal{L}_{\left(  2\right)  }  &  =4R^{\mu\nu}R_{\mu\nu}-R^{2}-R_{\alpha
\mu\beta\nu}R^{\alpha\mu\beta\nu},\\
\mathcal{L}_{\left(  3\right)  }  &  =R^{3}-12RR^{\mu\nu}R_{\mu\nu}%
+16R_{\mu\nu}R_{\text{ \ }\rho}^{\mu}R^{\nu\rho}+24R_{\alpha\beta}R_{\mu\nu
}R^{\alpha\mu\beta\nu}+3RR_{\alpha\mu\beta\nu}R^{\alpha\mu\beta\nu}\\
&  -24R_{\alpha\rho}R_{\text{ \ }\mu\beta\nu}^{\rho}R^{\alpha\mu\beta\nu
}+4R_{\alpha\mu\beta\nu}R_{\text{ \ }\rho\sigma}^{\alpha\mu}R^{\rho\sigma
\beta\nu}-8R_{\alpha\mu\beta\nu}R_{\text{ }\rho\text{ }\sigma}^{\alpha\text{
}\beta}R^{\mu\rho\nu\sigma},%
\end{align*}
where \cite{Decanini2007}%
\[
\int_{\Omega}\mathcal{L}_{\left(  2\right)  }\sqrt{|g|}d^{4}x=0\ \ \ \text{and
\ }\ \mathcal{L}_{\left(  3\right)  }=0\text{.}%
\]

Therefore, the higher-order action including terms up to mass dimension six
contains 11 independent terms and can be written in the following form:
\begin{align}
S  &  =\frac{M_{Pl}^{2}}{2}\int_{\Omega}\left[  R+\frac{R^{2}}{2\kappa_{0}%
}+\frac{\alpha_{0}}{3\kappa_{0}^{2}}R^{3}-\frac{\beta_{0}}{2\kappa_{0}^{2}%
}R\square R+\frac{\gamma_{0}}{3\kappa_{0}^{2}}RR^{\mu\nu}R_{\mu\nu}%
+\frac{\theta_{0}}{2\kappa_{0}}C^{\alpha\mu\beta\nu}C_{\alpha\mu\beta\nu
}\right. \nonumber\\
&  +\frac{\theta_{1}}{2\kappa_{0}^{2}}C^{\alpha\mu\beta\nu}\square
C_{\alpha\mu\beta\nu}+\frac{\theta_{2}}{3\kappa_{0}^{2}}RC^{\alpha\mu\beta\nu
}C_{\alpha\mu\beta\nu}+\frac{\theta_{3}}{3\kappa_{0}^{2}}R_{\alpha\rho
}C_{\text{ \ }\mu\beta\nu}^{\rho}C^{\alpha\mu\beta\nu}\label{eq:action_weyl}\\
&  \left.  +\frac{\theta_{4}}{3\kappa_{0}^{2}}C_{\alpha\mu\beta\nu}C_{\text{
\ \ }\rho\sigma}^{\alpha\mu}C^{\rho\sigma\beta\nu}+\frac{\theta_{5}}%
{3\kappa_{0}^{2}}C_{\alpha\mu\beta\nu}C_{\text{ \ }\rho\text{ \ }\sigma
}^{\alpha\text{ \ }\beta}C^{\mu\rho\nu\sigma}\right]  \sqrt{|g|}%
d^{4}x,\nonumber
\end{align}
where $C_{\alpha\mu\beta\nu}$ is the Weyl tensor, and the Greek letters denote
constants, with $\kappa_{0}$ having mass dimension two, while the remaining
coefficients are dimensionless. The use of the Weyl tensor instead of the
Riemann tensor is convenient due to the conformal invariance of
$C_{\alpha\mu\beta\nu}$ \cite{Wald}.\footnote{In four dimensions the Weyl
tensor can be written as
\[
C_{\alpha\mu\beta\nu}=R_{\alpha\mu\beta\nu}
-\tfrac{1}{2}\big(g_{\alpha\beta}R_{\nu\mu}-g_{\alpha\nu}R_{\beta\mu}
+g_{\mu\nu}R_{\beta\alpha}-g_{\mu\beta}R_{\nu\alpha}\big)
+\tfrac{R}{6}\,(g_{\alpha\mu}g_{\beta\nu}-g_{\alpha\beta}g_{\mu\nu})\,.
\]}

The next step is to modify the action in Eq.~(\ref{eq:action_weyl}) following
the procedure described in Ref.~\cite{Wands1994}, which is analogous to the
transition from $f(R)$ models to the Jordan frame
\cite{Faraoni2010,Tsujikawa2010,Odintsov2011}. To this end, we start by
introducing a new action of the form
\begin{equation}
S^{\prime}  =\frac{M_{Pl}^{2}}{2}\int_{\Omega}\left[  \chi+\frac{\chi^{2}%
}{2\kappa_{0}}+\frac{\alpha_{0}}{3\kappa_{0}^{2}}\chi^{3}-\frac{\beta_{0}%
}{2\kappa_{0}^{2}}\chi\chi_{1}+\frac{\gamma_{0}}{3\kappa_{0}^{2}}\chi
R^{\mu\nu}R_{\mu\nu}+\mathcal{L}_{C}\right.
 \left.  +\phi_{0}\left(  R-\chi\right)  +\phi_{1}\left(  \square R-\chi
_{1}\right)  \right]  \sqrt{|g|}d^{4}x, \label{eq: action_S}%
\end{equation}
where%
\begin{align}
\mathcal{L}_{C}  &  =\frac{\theta_{0}}{4\kappa_{0}}C^{\alpha\mu\beta\nu
}C_{\alpha\mu\beta\nu}+\frac{\theta_{1}}{2\kappa_{0}^{2}}C^{\alpha\mu\beta\nu
}\square C_{\alpha\mu\beta\nu}+\frac{\theta_{2}}{3\kappa_{0}^{2}}\chi
C^{\alpha\mu\beta\nu}C_{\alpha\mu\beta\nu} \nonumber \\
&  +\frac{\theta_{3}}{3\kappa_{0}^{2}}R_{\alpha\rho}C_{\text{ \ }\mu\beta\nu
}^{\rho}C^{\alpha\mu\beta\nu}+\frac{\theta_{4}}{3\kappa_{0}^{2}}C_{\alpha
\mu\beta\nu}C_{\text{ \ \ }\rho\sigma}^{\alpha\mu}C^{\rho\sigma\beta\nu}%
+\frac{\theta_{5}}{3\kappa_{0}^{2}}C_{\alpha\mu\beta\nu}C_{\text{ }\rho\text{
}\sigma}^{\alpha\text{ }\beta}C^{\mu\rho\nu\sigma}.
\end{align}
Extremizing this action with respect to the Lagrange multipliers $\phi_{0}$ and
$\phi_{1}$ yields $R=\chi$ and $\square R=\chi_{1}$, which demonstrates the
on-shell equivalence between $S$ and $S^{\prime}$.

The next step is to obtain the equations associated with the fields $\chi$
and $\chi_{1}$. Varying the action (\ref{eq: action_S}) with respect to $\chi$
and $\chi_{1}$ we obtain, respectively,
\begin{equation}
\phi_{0}=1+\frac{\chi}{\kappa_{0}}+\frac{\alpha_{0}}{\kappa_{0}^{2}}\chi
^{2}-\frac{\beta_{0}}{2\kappa_{0}^{2}}\chi_{1}+\frac{\gamma_{0}}{3\kappa
_{0}^{2}}R^{\mu\nu}R_{\mu\nu}+\frac{\theta_{2}}{3\kappa_{0}^{2}}C_{\alpha
\mu\beta\nu}C^{\alpha\mu\beta\nu}, \label{Aux1}%
\end{equation}
and
\begin{equation}
\phi_{1}=-\frac{\beta_{0}}{2\kappa_{0}^{2}}\chi.
\end{equation}
We then invert these equations to obtain $\chi=\chi(\phi_{1})$ and
$\chi_{1}=\chi_{1}(\phi_{0},\phi_{1})$, and substitute the result back into the
action (\ref{eq: action_S}). After integrating by parts, the term
$\phi_{1}\square R$, and performing some algebraic manipulations, we arrive at
\begin{equation}
S =\frac{M_{Pl}^{2}}{2}\int d^{4}x\sqrt{-g}\left[  \left(  \phi
_{0}+\square\phi_{1}\right)  R+\frac{2\kappa_{0}^{2}}{\beta_{0}}\phi
_{1}\left(  \phi_{0}-1\right)  \right.
 \left.  +\frac{2\kappa_{0}^{3}}{\beta_{0}^{2}}\phi_{1}^{2}-\frac
{8\kappa_{0}^{4}\alpha_{0}}{3\beta_{0}^{3}}\phi_{1}^{3}-\frac{2\gamma_{0}%
}{3\beta_{0}}\phi_{1}R_{\mu\nu}R^{\mu\nu}+\mathcal{L}_{C}\right]  .
\end{equation}
Finally, defining
\begin{equation}
\lambda\equiv-\frac{2\kappa_{0}}{\beta_{0}}\phi_{1}\text{ \ and \ }\phi
\equiv\phi_{0}+\square\phi_{1},%
\end{equation}
the action can be rewritten as
\begin{equation}
S=\frac{M_{Pl}^{2}}{2}\int_{\Omega}\left[  \phi R-\kappa_{0}\lambda\left(
\phi-1-\frac{\lambda}{2}-\frac{\alpha_{0}}{3}\lambda^{2}\right)  +\frac
{\beta_{0}}{2}\nabla^{\sigma}\lambda\nabla_{\sigma}\lambda+\frac{\gamma_{0}%
}{3\kappa_{0}}\lambda R_{\mu\nu}R^{\mu\nu}+\mathcal{L}_{C}\right]  \sqrt
{|g|}d^{4}x. \label{eq:action_Jordan}%
\end{equation}

The action (\ref{eq:action_Jordan}) represents the original higher-order model
in the Jordan frame. In this frame, the degrees of freedom are given by the
metric $g_{\mu\nu}$ and two dimensionless scalar fields, $\phi$ and $\lambda$.
Taking the variation of this action with respect to $\phi$, $\lambda$, and
$g_{\mu\nu}$, we obtain the following field equations:
\begin{equation}
R=\kappa_{0}\lambda, \label{eq:field_phi}%
\end{equation}%
\begin{equation}
\kappa_{0}\left(  \phi-1-\lambda\right)  -\kappa_{0}\alpha_{0}\lambda
+\beta_{0}\square\lambda-\frac{\gamma_{0}}{3\kappa_{0}}R^{\mu\nu}R_{\mu\nu
}-\frac{\theta_{2}}{3\kappa_{0}}C_{\alpha\mu\beta\nu}C^{\alpha\mu\beta\nu}=0,
\label{eq:field_lambda}%
\end{equation}
and
\begin{align}
\phi R_{\mu\nu} &+ g_{\mu\nu}\square\phi-\nabla_{\mu}\nabla_{\nu}\phi+\frac
{\beta_{0}}{2}\nabla_{\mu}\lambda\nabla_{\nu}\lambda+\frac{\gamma_{0}}%
{3\kappa_{0}}\left[  R_{\mu\nu}\square\lambda+2\nabla_{\alpha}\lambda
\nabla^{\alpha}R_{\mu\nu}+\lambda\square R_{\mu\nu}-\frac{1}{2}\nabla_{(\mu
}\lambda\nabla_{\nu)}R\right. \nonumber\\&
\left.  - \nabla_{(\mu}\left(  R_{\nu)}{}^{\alpha}\nabla_{\alpha}%
\lambda\right)  -\lambda\nabla_{\mu}\nabla_{\nu}R-2\lambda R_{\beta\mu
\alpha\nu}R^{\alpha\beta}+g_{\mu\nu}\left(  \nabla^{\alpha}R\nabla_{\alpha
}\lambda+R^{\alpha\beta}\nabla_{\alpha}\nabla_{\beta}\lambda+\frac{1}%
{2}\lambda\square R\right)  \right] \label{eq:field_metric}\\
&- \frac{1}{2}g_{\mu\nu}\left[  \phi R-\kappa_{0}\left(  \phi-1-\frac{\lambda
}{2}\right)  \lambda+\frac{\kappa_{0}\alpha_{0}}{3}\lambda^{3}+\frac{\beta
_{0}}{2}\nabla^{\alpha}\lambda\nabla_{\alpha}\lambda+\frac{\gamma_{0}}%
{3\kappa_{0}}\lambda R^{\alpha\beta}R_{\alpha\beta}\right]  +\mathcal{C}%
_{\mu\nu}=0,\nonumber
\end{align}
where
\begin{equation}
\mathcal{C}_{\mu\nu}=\frac{\delta}{\delta g^{\mu\nu}}\int_{\Omega}%
\mathcal{L}_{C}\sqrt{|g|}d^{4}x. \label{Tensor C_munu}%
\end{equation}
The above equations will provide the basis for the analysis of inflationary
cosmology in the following sections.

\subsection{Cosmological Background Equations}

To derive the field equations in the cosmological background, we start by
considering the flat FLRW metric:
\begin{equation}
ds^{2}=dt^{2}-a^{2}\left(  dx^{2}+dy^{2}+dz^{2}\right)  . \label{eq:flrw_flat}%
\end{equation}
From the coordinate transformation $dt=a\,d\eta$ we see that the FLRW line
element is conformally flat, i.e. $ds^{2}=a^{2}\eta_{\mu\nu}dx^{\mu}dx^{\nu}$.
On the other hand, since the Weyl tensor is invariant under conformal
transformations, it vanishes identically when evaluated in the FLRW geometry.
Therefore, any term in the action (\ref{eq:action_Jordan}) involving two Weyl
tensors does not contribute to the field equations. It greatly simplifies the
generalized Friedmann equations, as the term $\mathcal{C}_{\mu\nu}$ in
Eq.~(\ref{eq:field_metric}) vanishes identically.

Substituting the metric (\ref{eq:flrw_flat}) into the field equations
(\ref{eq:field_phi}), (\ref{eq:field_lambda}), and (\ref{eq:field_metric}), we
obtain after a lengthy calculation the following set of four equations:
\begin{align}
h_{t}-\frac{\lambda}{6}+2h^{2}  & =0,\label{eq:h_dot-2}\\
\beta_{0}\left(  \lambda_{tt}+3h\lambda_{t}\right)  -\left(  \phi
-1-\lambda-\alpha_{0}\lambda^{2}\right)  +4\gamma_{0}\left(  \frac{\lambda
^{2}}{36}-\frac{\lambda}{6}h^{2}+h^{4}\right)    & =0,\label{eq:lambda-2}\\
3h\phi_{t}+\frac{\beta_{0}}{4}\lambda_{t}^{2}+3h^{2}\phi-\frac{\lambda}%
{2}\left(  \phi-1-\frac{\lambda}{2}-\frac{\alpha_{0}}{3}\lambda^{2}\right)
+\frac{2\gamma_{0}}{3}\left(  2h\lambda\lambda_{t}-3h^{3}\lambda_{t}%
+2h^{2}\lambda^{2}-6h^{4}\lambda+\frac{\lambda^{3}}{12}\right)    &
=0,\label{eq:g00-2}\\
\phi_{tt}-h\phi_{t}+2\left(  \frac{\lambda}{6}-2h^{2}\right)  \phi-\frac
{\beta_{0}}{2}\lambda_{t}^{2}+\frac{2\gamma_{0}}{3}\left(  \frac{2}{3}%
\lambda\lambda_{tt}-h^{2}\lambda_{tt}+\frac{2}{3}\lambda_{t}^{2}%
-h\lambda\lambda_{t}+5h^{3}\lambda_{t}-\frac{8}{3}h^{2}\lambda^{2}%
+16h^{4}\lambda+\frac{\lambda^{3}}{9}\right)    & =0,\label{eq:gsub-2}%
\end{align}
where $h$ is the Hubble function and the subscript $t$ denotes derivatives with
respect to the dimensionless time variable, namely
\begin{equation}
h\equiv\frac{a_{t}}{a}\text{ \ \ and \ \ }g_{t}\equiv\frac{1}{\sqrt{\kappa
_{0}}}\frac{dg}{dt}. \label{eq:h_t-1}%
\end{equation}
The definitions in Eq.~(\ref{eq:h_t-1}) render the field equations
dimensionless, which is convenient for the numerical analysis of the phase
space.

In the form in which the four field equations have been presented, Eq.~(\ref{eq:h_dot-2})
has already been used in the other three to eliminate the time derivatives
involving $h$. Thus, the system is completely described by the remaining three
equations. Moreover, we note that Eq.~(\ref{eq:g00-2}) is a first-order
differential equation in $\lambda$ and $\phi$, while algebraic in $h$.
Therefore, one can solve this equation for
$h=h(\lambda,\lambda_{t},\phi,\phi_{t})$ and substitute the result into the
other two equations. In this way, we find that the background cosmology is
governed by the second-order equations (\ref{eq:lambda-2}) and
(\ref{eq:gsub-2}), which together define a four-dimensional autonomous
dynamical system in the variables $\lambda,\lambda_{t},\phi$, and $\phi_{t}$.

Due to the simplicity of the FLRW metric, the original action
(\ref{eq:action_weyl}) reduces to
\begin{equation}
S=\frac{M_{Pl}^{2}}{2}\int_{\Omega}\left[  R+\frac{R^{2}}{2\kappa_{0}}%
+\frac{\alpha_{0}}{3\kappa_{0}^{2}}R^{3}-\frac{\beta_{0}}{2\kappa_{0}^{2}%
}R\square R+\frac{\gamma_{0}}{3\kappa_{0}^{2}}RR^{\mu\nu}R_{\mu\nu}\right]
\sqrt{|g|}d^{4}x. \label{main action}%
\end{equation}
In the inflationary context of interest here, this action can be interpreted as
follows: the first two terms constitute the Starobinsky model, while the last
three represent corrections to it. Thus, the three mass-dimension-six terms
correspond to possible higher-energy corrections to Starobinsky inflation.
Among these, the first and the second, involving $\alpha_{0}$ and $\beta_{0}$,
have already been explored in the literature. In fact, inflationary cosmology
involving only the $R^{3}$ correction was discussed in
Refs.~\cite{Huang2014,Cheong2020,SSM2022}, while the Starobinsky model plus the
$R\square R$ term was analyzed in
Refs.~\cite{Gottlober1990,Lihoshi2011,Castellanos2018,CMP2019}. The
inflationary dynamics, including both terms, was studied in
Refs.~\cite{Maeda1990,Gottlober1991,SM2023}. Therefore, in the remainder of
this paper, we shall focus on investigating the influence of the third
mass-dimension-six term on Starobinsky inflation, namely the model
$R+R^{2}+RR_{\mu\nu}R^{\mu\nu}$.

\section{Model $R+R^{2}+RR_{\mu\nu}R^{\mu\nu}$}

The field equations of the model we intend to investigate are obtained by setting $\alpha_{0}=\beta_{0}=0$. In this case, Eqs. (\ref{eq:lambda-2}), (\ref{eq:g00-2}), and (\ref{eq:gsub-2}) simplify to
\begin{align}
\phi-1-\lambda-\gamma_{0}\left(  \frac{\lambda^{2}}{9}-\frac{2\lambda}{3}%
h^{2}+4h^{4}\right)    & =0,\label{eq:lambda-3-1}\\
3h\phi_{t}+3h^{2}\phi-\frac{\lambda}{2}\left(  \phi-1-\frac{\lambda}%
{2}\right)  +\frac{2}{3}\gamma_{0}\left(  2h\lambda\lambda_{t}-3h^{3}%
\lambda_{t}+2h^{2}\lambda^{2}-6h^{4}\lambda+\frac{\lambda^{3}}{12}\right)    &
=0,\label{eq:g00-3-1}\\
\phi_{tt}-h\phi_{t}+2\left(  \frac{\lambda}{6}-2h^{2}\right)  \phi+\frac{2}%
{3}\gamma_{0}\left(  \frac{2}{3}\lambda\lambda_{tt}-h^{2}\lambda_{tt}+\frac
{2}{3}\lambda_{t}^{2}-h\lambda\lambda_{t}+5h^{3}\lambda_{t}-\frac{8}{3}%
h^{2}\lambda^{2}+16h^{4}\lambda+\frac{\lambda^{3}}{9}\right)    &
=0.\label{eq:gsub-3-1}%
\end{align}
The main difference of this model with respect to the general case is that Eq. (\ref{eq:lambda-3-1}) now becomes an algebraic equation for $\lambda$. Thus, it is possible to solve it for $\lambda$ in terms of $h$ and $\phi$, so that $h$, originally a function of $\lambda,$ $\lambda_{t},$ $\phi,$ and $\phi_{t}$, now depends only on $\phi$ and $\phi_{t}$. Consequently, the system is entirely governed by the dynamics of the field $\phi$. Therefore, we obtain a two-dimensional dynamical system in the variables $\phi$ and $\phi_{t}$.

Solving Eq. (\ref{eq:lambda-3-1}) for $\lambda$, we obtain
\begin{equation}
\lambda=\frac{\sqrt{\left(  6\gamma_{0}h^{2}-9\right)  ^{2}+4\gamma_{0}\left[
9\left(  \phi-1\right)  -36\gamma_{0}h^{4}\right]  }+6\gamma_{0}h^{2}%
-9}{2\gamma_{0}}, \label{eq:lambda_phi}%
\end{equation}
where the choice of sign was made in order to recover the Starobinsky model in the limit $\gamma_{0}\rightarrow0$.

In principle, we can use (\ref{eq:lambda_phi}) to eliminate the dependence on $\lambda$ and $\lambda_{t}$ in Eq. (\ref{eq:g00-3-1}). However, the algebraic equation for $h$ that emerges does not admit an obvious analytical solution. Therefore, for the sake of simplicity, we assume that $\gamma_{0}$ is a small quantity so that the equations can be linearized in $\gamma_{0}$.\footnote{As we shall see later, consistency with observations requires $\left\vert \gamma_{0}\right\vert <10^{-3}$.} In this case, after a series of manipulations, we obtain
\begin{align}
h &  \approx h_{s}+\gamma_{0}h_{p},\label{eq:h_lin_Jordan}\\
\phi_{tt} &  \approx h_{s}\phi_{t}-\frac{1}{3}\phi\left(  \phi-1\right)
+4h_{s}^{2}\phi+\frac{2}{3}\gamma_{0}\left[  \frac{3}{2}\left(  \phi
_{t}+8h_{s}\phi\right)  h_{p}+\frac{\left(  3\phi+2\right)  \left(
\phi-1\right)  ^{2}}{18}\right.  \nonumber\\
&  \left.  -\frac{2}{3}\phi_{t}^{2}-2h_{s}^{4}\left(  5\phi-8\right)
-4h_{s}^{3}\phi_{t}-\frac{2}{3}h_{s}^{2}\left(  \phi+4\right)  \left(
\phi-1\right)  +\frac{1}{3}h_{s}\phi_{t}\left(  \phi-1\right)  \right]
,\label{eq:field_star_RRicci}%
\end{align}
where%
\begin{align}
h_{s}  & =\frac{\sqrt{9\phi_{t}^{2}+3\phi\left(  \phi-1\right)  ^{2}}%
-3\phi_{t}}{6\phi},\label{eq:h_s}\\
h_{p}  & =\frac{2}{3\phi_{t}+6h_{s}\phi}\left[  2h_{s}^{4}\left(
\phi-1\right)  +h_{s}^{3}\phi_{t}-\frac{2}{3}h_{s}^{2}\left(  \phi-1\right)
^{2}-\frac{2}{3}h_{s}\phi_{t}\left(  \phi-1\right)  -\frac{\left(
\phi-1\right)  ^{3}}{36}\right]  .\label{eq:h_p}%
\end{align}
Equations (\ref{eq:field_star_RRicci}) and (\ref{eq:h_lin_Jordan}) determine the cosmological dynamics in the Jordan frame in the approximation $\left\vert \gamma_{0}\right\vert \ll1$.

The next step is to rewrite the field equations in the Einstein frame. Starting from the transformations
\begin{equation}
\phi=e^{\chi}\text{ \ e \ }\tilde{g}_{\mu\nu}=e^{\chi}g_{\mu\nu},
\label{Trans_Frame_Eins}%
\end{equation}
where the tilde denotes quantities in the Einstein frame, we can write
\[
ds^{2}  =d\tilde{t}^{2}-\tilde{a}^{2}\left(  d\tilde{x}^{2}+d\tilde{y}%
^{2}+d\tilde{z}^{2}\right)  =e^{\chi}dt^{2}-e^{\chi}a^{2}\left(  dx^{2}%
+dy^{2}+dz^{2}\right)  \Rightarrow
d\tilde{t}\text{ }  =e^{\frac{\chi}{2}}dt\text{\ \ and \ \ }\tilde
{a}=e^{\frac{\chi}{2}}a.
\]
Furthermore, by defining the dimensionless derivative
\begin{equation}
f_{\tau}\equiv\frac{1}{\sqrt{\kappa_{0}}}\frac{df}{d\tilde{t}},
\label{Der t frame Einstein}%
\end{equation}
it is straightforward to show that $\tilde{h}$ is given by
\begin{equation}
\tilde{h}=e^{-\frac{\chi}{2}}h+\frac{\chi_{\tau}}{2}\text{\ \ \ where
\ }\tilde{h}\equiv\frac{\tilde{a}_{\tau}}{\tilde{a}}\text{.}
\label{h and h_til}%
\end{equation}
Thus, we can rewrite Eqs. (\ref{eq:field_star_RRicci}) and
(\ref{eq:h_lin_Jordan}) in terms of the new variables $\chi$ and $\tilde{h}$:
\begin{align}
\chi_{\tau\tau}  & \approx -3\tilde{h}_{s}\chi_{\tau}-\frac{1}{3}e^{-\chi
}\left(  1-e^{-\chi}\right)  +\gamma_{0}e^{\chi}P(\chi,\chi_{\tau
}),\label{Chi FE}\\
\tilde{h}  & \approx\frac{1}{2}\sqrt{\chi_{\tau}^{2}+\frac{1}{3}\left(
1-e^{-\chi}\right)  ^{2}}+\gamma_{0}e^{\chi}\tilde{h}_{p},\label{h til FE}%
\end{align}
where%
\begin{align}
\tilde{h}_{s}  &  =\frac{1}{2}\sqrt{\chi_{\tau}^{2}+\frac{1}{3}\left(
1-e^{-\chi}\right)  ^{2}},\label{h til s}\\
\tilde{h}_{p}  &  =\frac{1}{3\tilde{h}_{s}}\left\{  2\tilde{h}_{s}^{4}\left(
1-e^{-\chi}\right)  -\chi_{\tau}\tilde{h}_{s}^{3}\left(  3-4e^{-\chi}\right)
+\tilde{h}_{s}^{2}\left[  \frac{3}{2}\chi_{\tau}^{2}\left(  1-2e^{-\chi
}\right)  -\frac{2}{3}\left(  1-e^{-\chi}\right)  ^{2}\right]  \right.
\nonumber\\
&  \left.  -\chi_{\tau}\tilde{h}_{s}\left[  \frac{1}{4}\chi_{\tau}^{2}\left(
1-4e^{-\chi}\right)  +\frac{2}{3}e^{-\chi}\left(  1-e^{-\chi}\right)  \right]
+\frac{\chi_{\tau}^{2}}{2}\left[  \frac{1}{3}\left(  1-e^{-2\chi}\right)
-\frac{\chi_{\tau}^{2}}{4}e^{-\chi}\right]  -\frac{\left(  1-e^{-\chi}\right)
^{3}}{36}\right\}  , \label{h til p}%
\end{align}
and%
\begin{align}
P(\chi,\chi_{\tau})  &  =\frac{2}{3}\left\{  \frac{3}{2}\left(  8\tilde{h}%
_{s}-3\chi_{\tau}\right)  \tilde{h}_{p}+\frac{1}{3}\chi_{\tau}\tilde{h}%
_{s}\left[  6\chi_{\tau}^{2}\left(  1-4e^{-\chi}\right)  +\left(  3+8e^{-\chi
}\right)  \left(  1-e^{-\chi}\right)  \right]  \right. \nonumber\\
&  -\frac{1}{3}\tilde{h}_{s}^{2}\left[  2\left(  1+4e^{-\chi}\right)  \left(
1-e^{-\chi}\right)  +9\chi_{\tau}^{2}\left(  3-8e^{-\chi}\right)  \right]
+16\chi_{\tau}\tilde{h}_{s}^{3}\left(  1-2e^{-\chi}\right)  -2\tilde{h}%
_{s}^{4}\left(  5-8e^{-\chi}\right) \nonumber\\
&  \left.  +\frac{1}{18}\left(  3+2e^{-\chi}\right)  \left(  1-e^{-\chi
}\right)  ^{2}-\frac{\chi_{\tau}^{2}}{3}\left(  1+2e^{-\chi}\right)  \left(
1-e^{-\chi}\right)  -\frac{2}{3}\chi_{\tau}^{2}-\frac{\chi_{\tau}^{4}}%
{8}\left(  1-8e^{-\chi}\right)  \right\}  . \label{eq:P-g}%
\end{align}
Finally, we can combine Eqs. (\ref{Chi FE}) and (\ref{h til FE}) to obtain
\begin{equation}
\tilde{h}_{\tau}\approx-\frac{3}{4}\chi_{\tau}^{2}+\gamma_{0}\left[
\frac{P(\chi,\chi_{\tau})}{4\tilde{h}_{s}}\chi_{\tau}+\left(  e^{\chi}%
\tilde{h}_{p}\right)  _{\tau}\right]  . \label{h tal FE}%
\end{equation}
Equations (\ref{h til FE}) and (\ref{h tal FE}) represent the generalized Friedmann equations, which together with Eq. (\ref{Chi FE}) describe the dynamics of the cosmological background in the Einstein frame. Note that by setting $\gamma_{0}=0$, we recover the Starobinsky model in the same notation used in Refs. \cite{SSM2022,SM2023}. An important point to emphasize is that the structure of Eqs. (\ref{Chi FE}) and (\ref{h til FE}) shows that the corrections to the Starobinsky model are of order $\gamma_{0}e^{\chi}$.\footnote{As we will see in Sec. \ref{sec - SRI}, both $\left\vert P\left( \chi,\chi_{\tau
}\right) \right\vert $ and $\left\vert \tilde{h}{p}\right\vert $ are typically smaller than unity.} Therefore, the linearized equations provide good approximations for the model only when $\left\vert \gamma_{0}e^{\chi}\right\vert \ll1$.

Once the expressions in the Einstein frame are established, we can study the dynamical system of the present model. Defining $\Phi\equiv\chi_{\tau}$, we rewrite Eq. (\ref{Chi FE}) as a first-order autonomous system given by
\begin{equation}%
\begin{cases}
\chi_{\tau}=\Phi,\\
\Phi_{\tau}\approx-3\tilde{h}_{s}\Phi-\frac{1}{3}e^{-\chi}\left(  1-e^{-\chi
}\right)  +\gamma_{0}e^{\chi}P(\chi,\Phi).
\end{cases}
\label{eq:star_RRicciRicci_sist}%
\end{equation}
The critical points of this system are obtained from $\chi_{\tau}=\Phi_{\tau}=0$ and are given by
\begin{align}
P_{0}  &  :\left(  \chi_{0},\Phi_{0}\right)  =\left(  0,0\right)
,\label{P0}\\
P_{c}  &  :\left(  \chi_{c},\Phi_{c}\right)  \approx\left(  \ln\left(
\frac{6}{\sqrt{-17\gamma_{0}}}\right)  ,0\right)  , \label{Pc}%
\end{align}
where $P_{c}$ exists only when $\gamma_{0}<0$.

\begin{figure}[ht]
\centering
\includegraphics[width=0.49\linewidth]{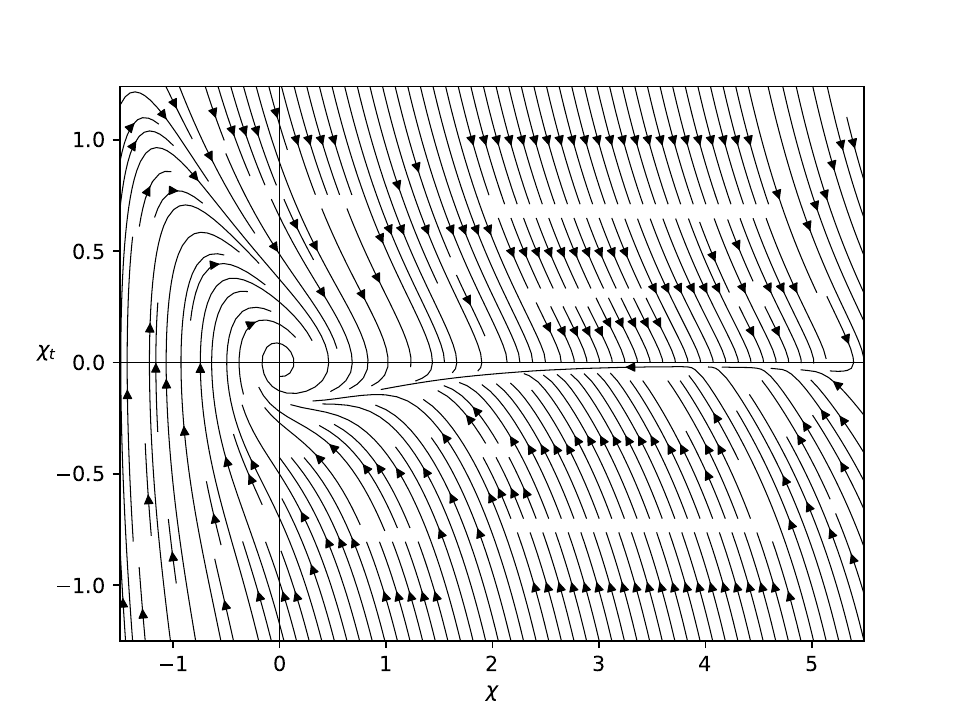}
\hfill
\includegraphics[width=0.49\linewidth]{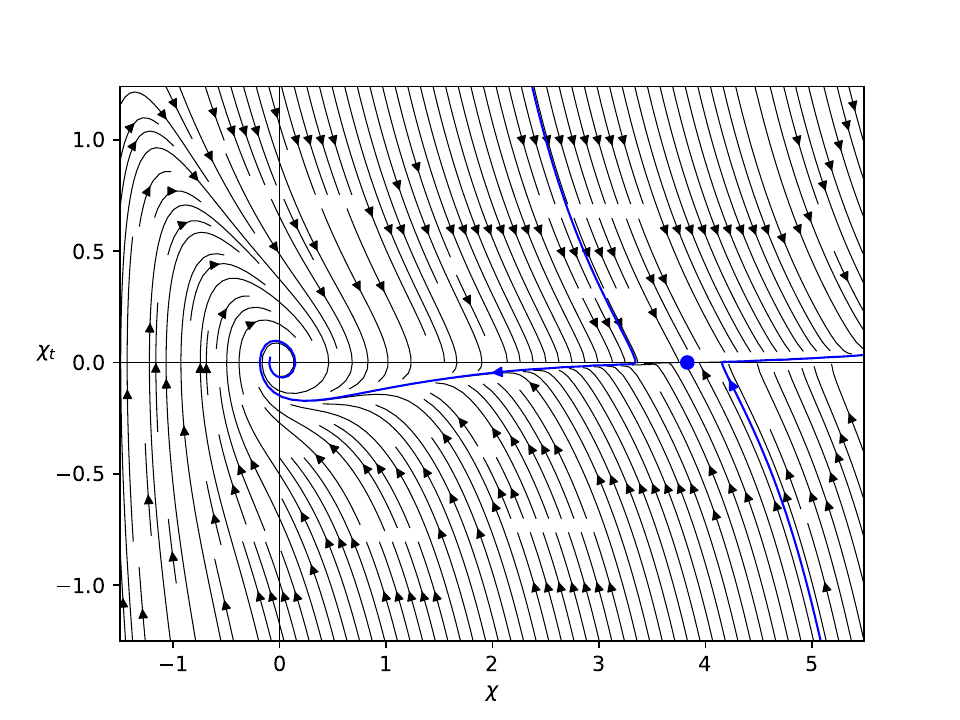}
\caption{\label{fig:phase_space} 
Phase spaces of the $R+R^{2}+RR_{\mu\nu}R^{\mu\nu}$ model
for $\gamma_{0}=10^{-3}$ (Left) and $\gamma_{0}=-10^{-3}$ (Right). The blue point in the second plot represents the critical point $P_{c}=\left(
3.83,0\right)  $.}
\end{figure}

The phase space associated with the system (\ref{eq:star_RRicciRicci_sist}) is shown in Fig. \ref{fig:phase_space}.

Both plots exhibit a behavior similar to the phase space of Starobinsky inflation, which features two main regions. The first is the horizontal attractor line associated with $\chi\gtrsim3$ and $\chi_{\tau}\approx0$, corresponding to a slow-roll inflationary dynamics (see Sec. \ref{sec - SRI}). The second is the region around the stable critical point $P_{0}=\left( 0,0\right)$, consistent with the reheating phase that connects the inflationary period to a hot Big Bang model.

Although both plots display a well-behaved inflationary structure for various sets of initial conditions, there is a fundamental difference between the two cases. For $\gamma_{0}>0$, any trajectory of the field $\chi$ that reaches the attractor line necessarily evolves toward the critical point $P_{0}$, leading to the end of inflation and the onset of reheating. In contrast, for $\gamma_{0}<0$, the outcome depends on the initial point at which the trajectory of $\chi$ intersects the attractor line. Suppose this point lies to the left of $P_{c}$, inflation proceeds in the usual way and evolves toward $P_{0}$. However, if the intersection occurs to the right of $P_{c}$, the trajectory evolves toward increasingly larger values of $\chi$, and the inflationary regime never ends. Therefore, for negative $\gamma_{0}$, there is a restriction on the set of initial conditions that lead to a consistent inflationary regime. Moreover, a similar behavior arises in the model involving the sixth-order mass correction $R^{3}$ \cite{Huang2014,Cheong2020,SSM2022}.

\subsection{Slow-roll inflation\label{sec - SRI}}

The slow-roll regime, associated with the horizontal attractor line in Fig. \ref{fig:phase_space}, occurs when $\chi_{\tau}\approx0$ and $\chi\gtrsim3$. To write the background equations in this regime, we define the fundamental slow-roll parameter $\delta\equiv e^{-\chi}$ and assume that the contribution coming from the additional $RR^{\mu\nu}R_{\mu\nu}$ term is, at most, of the same order as the Starobinsky term. Thus, from the structure of Eq. (\ref{Chi FE}), we have
\begin{equation}
\gamma_{0}e^{\chi}\sim e^{-\chi}\Rightarrow\gamma_{0}\lesssim\delta
^{2}\text{.} \label{gamma SR}%
\end{equation}
Hence, at leading order in slow-roll,
\[
\chi_{\tau}\sim\delta\Rightarrow\chi_{\tau\tau}\sim\delta^{2}\text{,}%
\]
which yields
\begin{equation}
\tilde{h}_{s}\approx\frac{1}{\sqrt{12}},\text{ \ }\tilde{h}_{p}\approx
-\frac{1}{12}\left(  \frac{5\sqrt{3}}{9}\right)  \text{ \ and \ }P(\chi
,\chi_{\tau})\approx-\frac{17}{108}. \label{Aprox SR}%
\end{equation}
Therefore, in the slow-roll approximation, the background equations (\ref{Chi FE}), (\ref{h til FE}), and (\ref{h tal FE}) are rewritten as
\begin{align}
\chi_{\tau}  &  \approx-\frac{2\sqrt{3}}{9}\delta\left(  1+\frac{17}{36}%
\gamma_{0}\delta^{-2}\right)  ,\label{Chi de tau}\\
\tilde{h}  &  \approx\frac{1}{\sqrt{12}}\left(  1-\delta-\frac{5}{18}%
\gamma_{0}\delta^{-1}\right)  ,\label{h til}\\
\tilde{h}_{\tau}  &  \approx-\left(  \frac{3}{4}\chi_{\tau}+\frac{\sqrt{3}}%
{8}\gamma_{0}\delta^{-1}\right)  \chi_{\tau}. \label{h tau anterior}%
\end{align}

From these three equations, it is straightforward to verify that the slow-roll dynamics results in an inflationary regime. Defining
\begin{equation}
\bar{\epsilon}\left(  \chi\right)  \equiv-\frac{\tilde{h}_{\tau}\left(
\chi\right)  }{\tilde{h}^{2}\left(  \chi\right)  }, \label{epsilon_bar}%
\end{equation}
we see that quasi-exponential expansion occurs when $\bar{\epsilon}\ll1$. Substituting expressions (\ref{Chi de tau}), (\ref{h til}), and (\ref{h tau anterior}) into Eq. (\ref{epsilon_bar}), and after linearizing in $\gamma_{0}$, we obtain
\begin{equation}
\bar{\epsilon}\approx\frac{4}{3}\delta^{2}\left(  1+\frac{7}{36}\gamma
_{0}\delta^{-2}\right)  . \label{epsilon_bar1}%
\end{equation}
Therefore, for $\chi\gtrsim3$ we have $\delta^{2}\ll1\Rightarrow\bar{\epsilon}\ll1$.

The next step is to redefine the scalar field in order to obtain a canonical kinetic term. Starting from Eq. (\ref{h tau anterior}) and considering the linearization in $\gamma_{0}$, we can write
\[
\tilde{h}_{\tau} \approx-\frac{3}{4}\left(  1-\frac{3}{8}\gamma
_{0}e^{2\chi}\right)  ^{2}\chi_{\tau}^{2}.
\]
Thus, by defining
\begin{equation}
\varphi_{\tau}\equiv\left(  1-\frac{3}{8}\gamma_{0}e^{2\chi}\right)
\chi_{\tau}, \label{Phi tau e Chi tau}%
\end{equation}
we can rewrite $\tilde{h}{\tau}$ as
\begin{equation}
\tilde{h}_{\tau}\approx-\frac{3}{4}\varphi_{\tau}^{2}. \label{h tau phi}%
\end{equation}
Integrating Eq. (\ref{Phi tau e Chi tau}) yields
\begin{equation}
\varphi\left(  \chi\right)  \approx\chi-\frac{3}{16}\gamma_{0}e^{2\chi
}\Rightarrow\chi\left(  \varphi\right)  \approx\varphi+\frac{3}{16}\gamma
_{0}e^{2\varphi}.\label{Chi de Phi}%
\end{equation}
Finally, substituting Eq. (\ref{Chi de Phi}) into Eqs. (\ref{Chi de tau}) and (\ref{h til}), we arrive at the expressions
\begin{align}
\tilde{h}^{2}  &  \approx\frac{1}{2}V\left(  \varphi\right)  ,\label{h phi}\\
3\tilde{h}\varphi_{\tau}  &  \approx-\frac{dV\left(  \varphi\right)
}{d\varphi}, \label{phi de tau}%
\end{align}
where%
\begin{equation}
V\left(  \varphi\right)  \approx\frac{1}{6}\left(  1-e^{-\varphi}-\frac
{13}{144}\gamma_{0}e^{\varphi}\right)  ^{2}, \label{Potential}%
\end{equation}
must be understood as valid under the condition of linearization in $\gamma_{0}$.

Equations (\ref{h tau phi}), (\ref{h phi}), and (\ref{phi de tau}) represent the cosmological background equations described by a canonical scalar field in the slow-roll approximation.\footnote{See the similarity with Eqs. (12), (19), and (20) of Ref. \cite{SSM2022}.}

The mapping $\chi\rightarrow\varphi$ makes it possible to compare the proposed model with CMB observations associated with the scalar spectral index $n_{s}$ and the tensor-to-scalar ratio $r$ \cite{Planck2020,BICEP3}. From the usual expressions \cite{Baumann}
\begin{equation}
n_{s}=1+\eta-2\epsilon\text{ \ \ and \ \ \ }r=16\epsilon, \label{Data}%
\end{equation}
where%
\begin{equation}
\epsilon\equiv-\frac{\tilde{h}_{\tau}}{\tilde{h}^{2}}\text{ \ and \ \ }%
\eta\equiv-\frac{1}{\tilde{h}}\frac{\epsilon_{\tau}}{\epsilon},
\label{SR parameters}%
\end{equation}
we can obtain an estimated constraint on the parameter $\gamma_{0}$.
Substituting Eqs. (\ref{h tau phi}) and (\ref{h phi}) into the definitions given in (\ref{SR parameters}), we find
\begin{align}
\epsilon &  \approx\frac{4}{3}e^{-2\varphi}\left(  1-\frac{13}{72}\gamma
_{0}e^{2\varphi}\right)  ,\label{epsilon phi}\\
\eta &  \approx-\frac{8}{3}e^{-\varphi}\left(  1+\frac{13}{144}\gamma
_{0}e^{2\varphi}\right)  . \label{eta phi}%
\end{align}

The final step is to obtain the number of e-folds $N=N\left(\varphi\right)$ and rewrite the slow-roll parameters in terms of $N$. Starting from the definition of the number of e-folds, one finds
\begin{equation}
N\equiv
{\displaystyle\int\limits_{t}^{t_{e}}}
Hdt=%
{\displaystyle\int\limits_{\varphi}^{\varphi_{e}}}
\frac{\tilde{h}}{\varphi_{\tau}}d\varphi\Rightarrow N\approx\frac{3}{4}\left(
e^{\varphi}+\frac{13}{432}\gamma_{0}e^{3\varphi}\right)  , \label{N}%
\end{equation}
where we assume $e^{\varphi}\gg e^{\varphi_{e}}$. The latter expression can be algebraically inverted by considering the linearization in $\gamma_{0}$. Thus, we obtain
\begin{equation}
e^{\varphi}\approx\frac{4}{3}N\left(  1-\frac{13}{243}\gamma_{0}N^{2}\right),
\label{phi de N}%
\end{equation}
and%
\begin{align}
\epsilon &  \approx\frac{3}{4N^{2}}\left(  1-\frac{52}{243}\gamma_{0}%
N^{2}\right)  ,\label{epsilon N}\\
\eta &  \approx-\frac{2}{N}\left(  1+\frac{52}{243}\gamma_{0}N^{2}\right)  .
\label{eta N}%
\end{align}
Therefore, at leading order in slow-roll, we can write the spectral index and the tensor-to-scalar ratio as
\begin{align}
n_{s}  &  \approx1-\frac{2}{N}\left(  1+\frac{52}{243}\gamma_{0}N^{2}\right)
,\label{ns}\\
r  &  \approx\frac{12}{N^{2}}\left(  1-\frac{52}{243}\gamma_{0}N^{2}\right)  .
\label{r}%
\end{align}

Figure \ref{fig:r_ns} shows the evolution of the parameter $\gamma_{0}$ in the $n_{s}\times r$ plane, together with observational data in blue \cite{BICEP3}:

\begin{figure}[ht]
\centering
\includegraphics[width=0.7\linewidth]{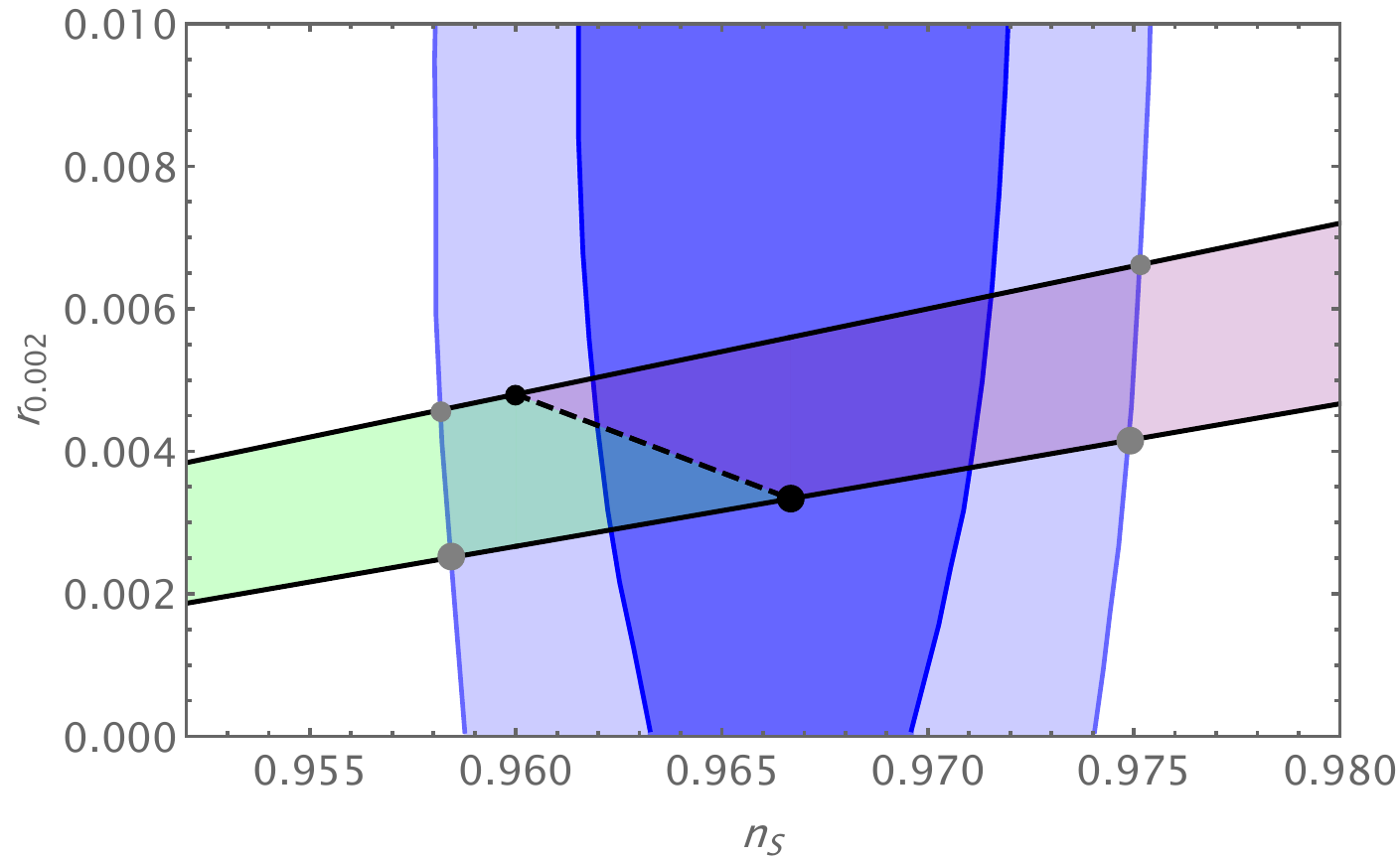}
\caption{\label{fig:r_ns} 
The blue contours correspond to $68\%$ and $95\%$ C.L.
constraints on $n_{s}\times r_{0.002}$ given by Planck plus BICEP3/Keck plus
BAO data \cite{BICEP3}. The black circles represent the Starobinsky model
($\gamma_{0}=0$) for $N=50$ (smaller one) and $N=60$ (bigger one). As
$\gamma_{0}$ increases (decreases), the curves move to the left (right) light
green (light purple) region. The grey circles represent the upper limits for
$\gamma_{0}$ associated with $95\%$ C.L.. Table \ref{tab:grey_circles} provides the $\gamma_{0}$
numerical values associated with the grey circles.}
\end{figure}

Taking into account the interval $50<N<60$, Fig. \ref{fig:r_ns} shows how the values of $n_{s}$ and $r$ evolve with the variation of $\gamma_{0}$. As $\gamma_{0}$ becomes more negative, the values of $n_{s}$ and $r$ move to the right and upward (light purple region), increasing both the tensor-to-scalar ratio and the scalar spectral index. Conversely, as $\gamma_{0}$ increases in the positive direction, the points shift to the left and downward (light green region), decreasing the tensor-to-scalar ratio and the spectral index. The gray circles on the left and right correspond to the $95\%$ C.L. limits, with the smaller and larger circles associated with $N=50$ and $N=60$, respectively. Table \ref{tab:grey_circles} summarizes the values of $\gamma_{0}$, $n_{s}$, and $r$ for each of the four circles:

\begin{table}[ht]
\caption{\label{tab:grey_circles} Limiting values at 95\% C.L. associated with the gray circles in Fig. \ref{fig:r_ns}. Note that the largest value of $\gamma_{0}$ still satisfies the condition $\left\vert \gamma_{0}\right\vert <10^{-3}$.}
\begin{ruledtabular}
\begin{tabular}{ccccc}
Grey circles & $N$ & $\gamma_{0}$ & $n_{s}$ & $r$ \\
\hline
smaller left  & $50$ & $9\times10^{-5}$  & $0.9582$ & $0.0046$ \\
bigger left   & $60$ & $3\times10^{-4}$  & $0.9584$ & $0.0025$ \\
smaller right & $50$ & $-7\times10^{-4}$ & $0.9752$ & $0.0066$ \\
bigger right  & $60$ & $-3\times10^{-4}$ & $0.9749$ & $0.0042$ \\
\end{tabular}
\end{ruledtabular}
\end{table}

Finally, we must verify whether the results obtained from Eqs. (\ref{ns}) and (\ref{r}) are consistent with the approximations performed.

The first approximation concerns the linearization of the original equations (\ref{eq:lambda-3-1}), (\ref{eq:g00-3-1}), and (\ref{eq:gsub-3-1}). As discussed in the text below Eq. (\ref{h tal FE}), the validity of the linearized equations requires not only $\left\vert \gamma_{0}\right\vert \ll1$, but also $\left\vert \gamma_{0}e^{\chi}\right\vert \ll1$. Thus, from Eqs. (\ref{Chi de Phi}) and (\ref{phi de N}) we can write
\begin{equation}
\left\vert \gamma_{0}e^{\chi}\right\vert \ll1\Rightarrow\left\vert \gamma
_{0}e^{\varphi}\right\vert \ll1\Rightarrow\left\vert \frac{4}{3}\gamma
_{0}N\right\vert \ll1. \label{Con1}%
\end{equation}
Therefore, even in the worst-case scenario ($N=50$ and $\gamma_{0}=-7\times10^{-4}$), we obtain $\left\vert \gamma_{0}e^{\chi}\right\vert \lesssim0.05$, which is consistent with condition (\ref{Con1}).

The second approximation is related to the assumption that the $RR^{\mu\nu}R_{\mu\nu}$ term is mostly of the same order as the Starobinsky term. In Eqs. (\ref{ns}) and (\ref{r}), this assumption implies
\begin{equation}
\left\vert \frac{52}{243}\gamma_{0}N^{2}\right\vert \lesssim1\Rightarrow
\left\vert 0.214\gamma_{0}N^{2}\right\vert \lesssim1. \label{Con2}%
\end{equation}
Thus, considering again the worst case in Table \ref{tab:grey_circles}, we find $\left\vert 0.214\gamma_{0}N^{2}\right\vert \approx0.37$, which is therefore consistent with condition (\ref{Con2}).

\section{Final Remarks}

In this work, we constructed a higher-order gravity model containing all corrections up to second order in General Relativity (mass dimension six). The general formulation was obtained in the Jordan frame, with an explicit derivation of the associated field equations. We then analyzed the field equations in the FLRW cosmological background, revealing that the resulting system has a four-dimensional autonomous dynamical system structure. Subsequently, we derived the equations corresponding to the specific $R+R^{2}+RR_{\mu\nu}R^{\mu\nu}$ model in FLRW and, afterwards, obtained the linearized field equations with respect to the parameter $\gamma_{0}$.

The analysis of the resulting dynamics showed that the phase space of the linearized system — Eq. (\ref{eq:star_RRicciRicci_sist}) — exhibits the same structure as the Starobinsky model, with an attractor line associated with the slow-roll regime and a stable critical point corresponding to the end of inflation. A detailed investigation of this regime confirmed that the dynamics lead to a consistent inflationary period. In addition, we computed the model predictions for the scalar spectral index $n_{s}$ and the tensor-to-scalar ratio $r$, obtaining analytical expressions in Eqs. (\ref{ns}) and (\ref{r}). A direct comparison with the most recent data from the Planck collaboration, complemented by BICEP/Keck and BAO, shows that the model reproduces the observationally allowed region in the $\left( n_{s},r\right)$ plane. We also found that consistency constraints impose the restriction $\left\vert \gamma_{0}\right\vert \lesssim10^{-3}$, so that within this interval, higher-order corrections maintain the successful fit characteristic of the Starobinsky model.

Corrections to the Starobinsky inflationary model are particularly relevant in light of the increasingly stringent constraints imposed by current observations, e.g., the ACT and SPT collaborations \cite{ACT2025,SPT2025}, as well as by forthcoming measurements from next-generation experiments such as the Simons Observatory, CMB-S4, and the LiteBIRD satellite \cite{SO2024,S42022,LiteBird2024}. For instance, recent results suggest that the combined data from Planck, ACT, and DESI are in tension with the Starobinsky model in the usual interval $50<N<60$ at the $95\%$ C.L. \cite{ACT2025}. However, the extension proposed in this work restores full consistency with the observations by allowing negative values of $\gamma_{0}$, without the need to adjust the number of e-folds. As future perspectives, we emphasize the relevance of refining the $R+R^{2}+RR_{\mu\nu}R^{\mu\nu}$ model by considering the non-linear regime in $\gamma_{0}$ and performing an explicit perturbative analysis. These theoretical developments, together with forthcoming observational data, will be important to assess the viability of the model more accurately.

\begin{acknowledgments}
C. M. G. R. Morais thanks CAPES/UFRN-RN (Brazil) for financial support and L. G.
Medeiros acknowledges CNPq-Brazil (Grant No. 307901/2022-0) for partial financial support.
\end{acknowledgments}


\bibliography{referencias}

\end{document}